\documentclass[letter]{amsart}

\usepackage{amssymb}
\usepackage{amsfonts}
\usepackage{graphicx}

\newtheorem{thm}{Theorem}
\newtheorem{prob}{Problem}
\newtheorem{lem}{Lemma}
\newtheorem{ex}{Example}
\newcommand{\FIR}{{\mathcal F}}

\newcommand{\C}{{\mathbb C}}
\newcommand{\D}{{\mathbb D}}

\newcommand{\abcd}[4]{\left[\begin{array}{c|c}#1&#2\\\hline
			    #3&#4\end{array}\right]}

\newcommand{\real}{{\mathbb R}}

\renewcommand{\vec}[1]{#1}

\newcommand{\low}{{\mathrm{low}}}

\title[Min-Max Design of FIR Digital Filters]{Min-Max Design of FIR Digital Filters by Semidefinite Programming%
\footnote{Applications of Digital Signal Processing, pp.~193-210, 
InTech, 2011.}
}
\author[M. Nagahara]{Masaaki Nagahara}
\address{M.~Nagahara is with 
Graduate School of Informatics, Kyoto University,
Kyoto 606-8501, JAPAN. (nagahara@ieee.org).}

\begin{document}

\maketitle

\begin{abstract}
In this article we consider two problems:
FIR (Finite Impulse Response) approximation of IIR (Infinite Impulse Response)
filters and inverse FIR filtering of FIR or IIR filters.
By means of Kalman-Yakubovich-Popov (KYP) lemma and its generalization (GKYP),
the problems are reduced to semidefinite programming described in linear matrix inequalities (LMIs).
MATLAB codes for these design methods are given.
An design example shows the effectiveness of these methods.
\end{abstract}

\section{Introduction}
\label{sec:intro}
{\it Robustness} is a fundamental issue in signal processing;
unmodeled dynamics and unexpected noise in systems and signals
are inevitable in designing systems and signals.
Against such uncertainties, {\em min-max optimization},
or {\em worst case optimization} is a powerful tool.
In this light,
we propose an efficient design method
of FIR (finite impulse response) digital filters
for approximating and inverting given digital filters.
The design is formulated by {\em min-max optimization} in the frequency domain.
More precisely, we design an FIR filter which
minimizes the maximum gain of the frequency response of an error system.

This design has a direct relation with {\em $H^\infty$ optimization}
\cite{Fra}.
Since the space $H^\infty$ is not a Hilbert space,
the familiar projection method cannot be applied.
However, many studies have been made on the $H^\infty$ optimization,
and nowadays the optimal solution to the $H^\infty$ problem
is deeply analysed and 
can be easily obtained by numerical computation.
Moreover, as an extension of $H^\infty$ optimization,
a min-max optimization on a {\em finite} frequency interval
has been proposed recently \cite{IwaHar05}.
In both optimization, the {\em Kalman-Yakubovich-Popov (KYP) lemma} \cite{And67,Ran96,TuqVai98} and
the {\em (generalized) KYP lemma} \cite{IwaHar05}
give an easy and fast way of numerical computation;
{\em semidefinite programming} \cite{BoyVan}.
Semidefinite programming can be efficiently solved by numerical optimization
softwares.

In this article, we consider two fundamental problems of signal processing:
FIR approximation of IIR (infinite impulse response) filters and inverse FIR filtering of FIR/IIR filters.
Each problems are formulated in two types of optimization:
$H^\infty$ optimization and finite-frequency min-max one.
These problems are reduced to semidefinite programming in a similar way.
For this, we introduce state-space representation.
Semidefinite programming is obtained by the generalized KYP lemma.
We will give MATLAB codes for the proposed design,
and will show design examples.

\section{Preliminaries}
\label{sec:preliminaries}

In this article, we frequently use notations in control systems.
For readers who are not familiar to these,
we here recall basic notations and facts of control systems used throughout the article.
We also show MATLAB codes for better understanding.

Let us begin with a linear system ${\mathcal G}$ represented in the following 
{\em state-space equation} or {\em state-space representation} \cite{Rug}:
\begin{equation}
{\mathcal G}:\left\{~~
\begin{split}
x[k+1] &= Ax[k] + Bu[k],\\ y[k] &= Cx[k] + Du[k], ~~ k=0,1,2,\ldots.
\end{split}
\right.
\label{eq:state-space}
\end{equation}
The nonnegative number $k$ denotes the time index.
The vector $x[k]\in\real^n$ is called the state vector,
$u[k]\in\real$ is the input and $y[k]\in\real$ is the output of the system ${\mathcal G}$.
The matrices $A\in\real^{n\times n}$, $B\in\real^{n\times 1}$, $C\in\real^{1\times n}$, and $D\in\real$
are assumed to be static, that is, independent of the time index $k$.
Then the {\em transfer function} $G(z)$ of the system ${\mathcal G}$ is defined by
\[
 G(z) := C(zI-A)^{-1}B + D.
\]
The transfer function $G(z)$ is a rational function of $z$ of the form
\begin{equation}
 G(z) = \frac{b_nz^n+b_{n-1}z^{n-1}+\cdots+b_1z+b_0}{z^n+a_{n-1}z^{n-1}+\cdots+a_1z+a_0}.
 \label{eq:transfer-function}
\end{equation}
Note that $G(z)$ is the $Z$-transform of the impulse response $\{g[k]\}_{k=0}^\infty$ of the system ${\mathcal G}$ with
the initial state $x[0]=0$,
that is,
\[
 G(z) = \sum_{k=0}^\infty g[k] z^{-k} = D+\sum_{k=1}^\infty CA^{k-1}Bz^{-k}.
\]
To convert a state-space equation to its transfer function,
one can use the above equations or the MATLAB command \verb=tf=.
On the other hand, to convert a transfer function to a state-space equation,
one can use realization theory which provides methods to derive the 
state space matrices from a given transfer function
\cite{Rug}.
An easy way to obtain the matrices is to use
MATLAB or Scilab with the command \verb=ss=.
\vspace{5mm}
\begin{ex}
\label{ex:ss-matlab}
We here show an example of MATLAB commands.
First, we define state-space matrices:
\begin{verbatim}
>A=[0,1;-1,-2]; B=[0;1]; C=[1,1]; D=0;
>G=ss(A,B,C,D,1);
\end{verbatim}
This defines a state-space (ss) representation of ${\mathcal G}$ with the state-space matrices
\[
 A=\begin{bmatrix}0&1\\-1&-2\end{bmatrix},~B=\begin{bmatrix}0\\1\end{bmatrix},~C=\begin{bmatrix}1&1\end{bmatrix},~D=0.
\]
The last argument \verb=1= of \verb=ss= sets the sampling period to be 1.
\end{ex}
To obtain the transfer function $G(z)=C(zI-A)^{-1}B+D$, we can use the command \verb=tf=
\begin{verbatim}
>> tf(G)

Transfer function:
    z + 1
-------------
z^2 + 2 z + 1
 
Sampling time (seconds): 1
\end{verbatim}
On the other hand, suppose that we have a transfer function at first:
\begin{verbatim}
>> z=tf('z',1);
>> Gz=(z^2+2*z+1)/(z^2+0.5*z+1);
\end{verbatim}
The first command defines the variable $z$ of $Z$-transform with sampling period 1,
and the second command defines the following transfer function:
\[
 G(z) = \frac{z^2+2z+1}{z^2+0.5z+1}.
\]
To convert this to state-space matrices $A$, $B$, $C$, and $D$,
use the command \verb=ss= as follows:
\begin{verbatim}
>> ss(Gz)
 
a = 
         x1    x2
   x1  -0.5    -1
   x2     1     0
 
b = 
       u1
   x1   1
   x2   0
 
c = 
        x1   x2
   y1  1.5    0
 
d = 
       u1
   y1   1
 
Sampling time (seconds): 1
Discrete-time model.
\end{verbatim}
These outputs shows that the state-space matrices are given by
\[
 A=\begin{bmatrix}-0.5&-1\\1&0\end{bmatrix},~B=\begin{bmatrix}1\\0\end{bmatrix},~C=\begin{bmatrix}1.5&0\end{bmatrix},~D=1,
\]
with sampling time 1.
\hfill$\Box$

Note that the state-space representation in Example \ref{ex:ss-matlab}
is {\em minimal} in that the state-space model describes the same input/output
behavior with the minimum number of states.
Such a system is called {\em minimal realization} \cite{Rug}.

We then introduce a useful notation, called {\em packed notation} \cite{Vid88}, 
describing the transfer function from state-space matrices as
\[
 G(z) = C(zI-A)^{-1}B + D =: \abcd{A}{B}{C}{D}\!(z).
\]
By the packed notation,
the following formulae are often used in this article:
\begin{align}
 \abcd{A_1}{B_1}{C_1}{D_1}\times\abcd{A_2}{B_2}{C_2}{D_2}
 &= 
 \left[\begin{array}{cc|c}
  A_2 & 0 & B_2\\
  B_1C_2 & A_1 & B_1D_2\\\hline
  D_1C_2 & C_1 & D_1D_2\\
 \end{array}\right],\label{eq:mmult}\\
 \abcd{A_1}{B_1}{C_1}{D_1}\pm\abcd{A_2}{B_2}{C_2}{D_2}
 &= 
 \left[\begin{array}{cc|c}
  A_1 & 0 & B_1\\
  0 & A_2 & \pm B_2\\\hline
  C_1 & C_2 & D_1\pm D_2\\
 \end{array}\right].\label{eq:pm}
\end{align}

Next, we define {\em stability} of linear systems.
The state-space system ${\mathcal G}$ in (\ref{eq:state-space})
is said to be {\em stable} if the eigenvalues $\lambda_1,\ldots,\lambda_n$ of the matrix $A$
lie in the open unit circle $\D=\left\{z\in\C:\left|z\right|<1\right\}$.
Assume that the transfer function $G(z)$ is irreducible.
Then ${\mathcal G}$ is stable if and only if the poles of the transfer function $G(z)$ lie in $\D$.
To compute the eigenvalues of $A$ in MATLAB, use the command \verb=eig(A)=,
and for the poles of $G(z)$ use \verb=pole(Gz)=.

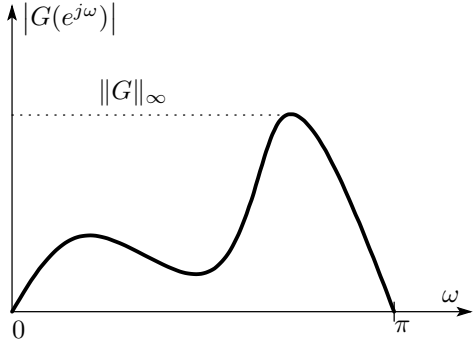
\begin{figure}[tbp]
\begin{center}
\unitlength 0.1in
\begin{picture}( 24.0500, 16.7000)(  3.9500,-18.5000)
%
\special{pn 8}%
\special{pa 400 1800}%
\special{pa 400 200}%
\special{fp}%
\special{sh 1}%
\special{pa 400 200}%
\special{pa 380 268}%
\special{pa 400 254}%
\special{pa 420 268}%
\special{pa 400 200}%
\special{fp}%
%
\special{pn 8}%
\special{pa 400 1800}%
\special{pa 2800 1800}%
\special{fp}%
\special{sh 1}%
\special{pa 2800 1800}%
\special{pa 2734 1780}%
\special{pa 2748 1800}%
\special{pa 2734 1820}%
\special{pa 2800 1800}%
\special{fp}%
\put(4.0000,-18.5000){\makebox(0,0)[lt]{$0$}}%
\put(24.0000,-18.5000){\makebox(0,0)[lt]{$\pi$}}%
%
\special{pn 8}%
\special{pa 2400 1750}%
\special{pa 2400 1850}%
\special{fp}%
%
\special{pn 20}%
\special{pa 400 1800}%
\special{pa 420 1766}%
\special{pa 440 1734}%
\special{pa 460 1700}%
\special{pa 480 1668}%
\special{pa 500 1636}%
\special{pa 522 1606}%
\special{pa 542 1576}%
\special{pa 564 1550}%
\special{pa 586 1524}%
\special{pa 608 1500}%
\special{pa 630 1478}%
\special{pa 654 1458}%
\special{pa 678 1440}%
\special{pa 702 1426}%
\special{pa 728 1414}%
\special{pa 754 1406}%
\special{pa 782 1402}%
\special{pa 810 1400}%
\special{pa 838 1402}%
\special{pa 868 1408}%
\special{pa 900 1416}%
\special{pa 930 1428}%
\special{pa 962 1440}%
\special{pa 994 1454}%
\special{pa 1026 1470}%
\special{pa 1058 1486}%
\special{pa 1092 1504}%
\special{pa 1124 1520}%
\special{pa 1156 1536}%
\special{pa 1188 1552}%
\special{pa 1218 1566}%
\special{pa 1248 1578}%
\special{pa 1278 1590}%
\special{pa 1308 1598}%
\special{pa 1336 1602}%
\special{pa 1364 1604}%
\special{pa 1390 1602}%
\special{pa 1414 1596}%
\special{pa 1438 1586}%
\special{pa 1460 1574}%
\special{pa 1480 1556}%
\special{pa 1500 1536}%
\special{pa 1520 1512}%
\special{pa 1538 1486}%
\special{pa 1554 1458}%
\special{pa 1570 1426}%
\special{pa 1586 1394}%
\special{pa 1600 1360}%
\special{pa 1614 1324}%
\special{pa 1628 1288}%
\special{pa 1640 1250}%
\special{pa 1654 1214}%
\special{pa 1666 1176}%
\special{pa 1676 1138}%
\special{pa 1688 1100}%
\special{pa 1698 1064}%
\special{pa 1710 1028}%
\special{pa 1720 992}%
\special{pa 1730 960}%
\special{pa 1742 928}%
\special{pa 1752 898}%
\special{pa 1762 872}%
\special{pa 1774 848}%
\special{pa 1784 826}%
\special{pa 1796 806}%
\special{pa 1808 792}%
\special{pa 1820 780}%
\special{pa 1832 772}%
\special{pa 1846 766}%
\special{pa 1858 766}%
\special{pa 1872 766}%
\special{pa 1886 770}%
\special{pa 1900 778}%
\special{pa 1914 788}%
\special{pa 1930 800}%
\special{pa 1944 816}%
\special{pa 1960 834}%
\special{pa 1976 854}%
\special{pa 1992 876}%
\special{pa 2008 900}%
\special{pa 2024 928}%
\special{pa 2042 956}%
\special{pa 2058 988}%
\special{pa 2074 1020}%
\special{pa 2092 1054}%
\special{pa 2110 1090}%
\special{pa 2128 1128}%
\special{pa 2144 1166}%
\special{pa 2162 1206}%
\special{pa 2180 1248}%
\special{pa 2198 1292}%
\special{pa 2218 1334}%
\special{pa 2236 1380}%
\special{pa 2254 1426}%
\special{pa 2272 1472}%
\special{pa 2292 1518}%
\special{pa 2310 1566}%
\special{pa 2328 1614}%
\special{pa 2348 1662}%
\special{pa 2366 1710}%
\special{pa 2384 1760}%
\special{pa 2400 1800}%
\special{sp}%
%
\special{pn 8}%
\special{pa 400 770}%
\special{pa 1850 770}%
\special{dt 0.045}%
\put(8.5000,-7.2000){\makebox(0,0)[lb]{$\|G\|_\infty$}}%
\put(26.5000,-17.5000){\makebox(0,0)[lb]{$\omega$}}%
\put(4.5000,-3.5000){\makebox(0,0)[lb]{$\left|G(e^{j\omega})\right|$}}%
\end{picture}%
\end{center}
\caption{The $H^\infty$ norm $\|G\|_\infty$ of $G(z)$ is the maximum value of the frequency response gain $\left|G(e^{j\omega})\right|$.}
\label{fig:hinf}
\end{figure}

The $H^\infty$ norm is the fundamental tool in this article.
The {\em $H^\infty$ norm} of a stable transfer function $G(z)$ is defined by
\[
 \|G\|_\infty := \max_{\omega\in[0,\pi]}\left|G(e^{j\omega})\right|.
\]
This is the maximum gain of the frequency response $G(e^{j\omega})$ of ${\mathcal G}$
as shown in Fig.~\ref{fig:hinf}.
The MATLAB code to compute the $H^\infty$ norm of a transfer function is given as follows:
\begin{verbatim}
>> z=tf('z',1);
>> Gz=(z-1)/(z^2-0.5*z);
>> norm(Gz,inf)

ans =

    1.3333
\end{verbatim}
This result shows that for the stable transfer function
\[
 G(z) = \frac{z-1}{z^2-0.5z},
\]
the $H^\infty$ norm is given by $\|G\|_\infty\approx 1.3333$.

$H^\infty$ control or $H^\infty$ optimization is thus minimization of the maximum value of a transfer function.
This leads to robustness against uncertainty in the frequency domain.
Moreover, it is known that the $H^\infty$ norm of a transfer function $G(z)$ is equivalent to the
{\em $\ell^2$-induced norm} of ${\mathcal G}$, that is,
\[
 \|G\|_\infty = \|{\mathcal G}\| := \sup_{\begin{subarray}{c} u\in \ell^2\\ u\neq 0\end{subarray}}\frac{\|{\mathcal G}u\|_2}{\|u\|_2},
\]
where $\|u\|_2$ is the $\ell^2$ norm of $u$:
\[
 \|u\|_2 := \left(\sum_{n=0}^\infty \left|u[k]\right|^2\right)^{1/2}.
\]
The $H^\infty$ norm optimization is minimization of the system gain when the worst case input is applied.
This fact implies that the $H^\infty$ norm optimization leads to robustness against uncertainty in input signals.

\section{$H^\infty$ Design Problems of FIR Digital Filters}
\label{sec:prob}

In this section, we consider two fundamental problems in signal processing:
filter approximation and inverse filtering.
The problems are formulated as $H^\infty$ optimization by using the $H^\infty$ norm
mentioned in the previous section.

\subsection{FIR approximation of IIR filters}
\label{subsec:prob:hinf-approximation}

The first problem we consider is {\em approximation}.
In signal processing, there are a number of design methods for
IIR (infinite impulse response) filters, e.g.,
Butterworth, Chebyshev, Elliptic, and so on \cite{OppSch}.
In general, to achieve a given characteristic, IIR filters require fewer memory elements, i.e.,
$z^{-1}$, than FIR (finite impulse response) filters.
However, IIR filters may have a problem of instability since they have feedbacks in their circuits,
and hence, we prefer an FIR filter to an IIR one in implementation.
For this reason, we employ FIR approximation of a given IIR filter.
This problem has been widely studied \cite{OppSch}.
Many of them are formulated by $H^2$ optimization; they aim at minimizing
the average error between a given IIR filter and the FIR filter to be designed.
This optimal filter works well {\em averagely}, but in the worst case, the filter
may lead a large error.
To guarantee the worst case performance,
$H^\infty$ optimization is applied to this problem \cite{YamAndNagKoy03}.
The problem is formulated as follows:

\begin{prob}[FIR approximation of IIR filters]
\label{prob:approximation}
Given an IIR filter $P(z)$,
find an FIR (finite impulse response) filter $Q(z)$
which minimizes
\[
\left\| (P-Q)W \right\|_\infty = \max_{\omega\in[0,\pi]} \left|\left(P(e^{j\omega})-Q(e^{j\omega})\right)W(e^{j\omega})\right|,
\]
where $W$ is a given stable weighting function.
\end{prob}

The procedure to solve this problem is shown in Section \ref{sec:kyp}.

\subsection{Inverse filtering}
\label{subsec:prob:hinf-inverse}

{\em Inverse filtering}, or {\em deconvolution} is another fundamental issue in signal processing.
This problem arises for example in direct-filter design
in spline interpolation \cite{NagYam11}.

Suppose a filter $P(z)$ is given.
Symbolically, the inverse filter of $P(z)$ is $P(z)^{-1}$.
However, real design is not that easy.
\begin{ex}
Suppose $P(z)$ is given by
\[
 P(z) = \frac{z+0.5}{z-0.5}.
\]
Then, the inverse $Q(z):=P(z)^{-1}$ becomes
\[
 Q(z) = P(z)^{-1} = \frac{z-0.5}{z+0.5},
\]
which is stable and causal.
Then suppose
\[
 P(z) = \frac{z-2}{z-0.5},
\]
then the inverse is
\[
 Q(z) = P(z)^{-1} = \frac{z-0.5}{z-2}.
\]
This has the pole at $\left|z\right|>1$, and hence the inverse filter is
unstable.
On the other hand, suppose
\[
 P(z) = \frac{1}{z-0.5},
\]
then the inverse is
\[
 Q(z) = P(z)^{-1} = z-0.5,
\]
which is noncausal.
\end{ex}

By these examples, the inverse filter $P(z)^{-1}$ may unstable or noncausal.
Unstable or noncausal filters are difficult to implement in real digital device,
and hence we adopt approximation technique;
we design an FIR digital filter $Q(z)$ such that $Q(z)P(z)\approx 1$.
Since FIR filters are always stable and causal, this is a realistic way to design an inverse filter.
Our problem is now formulated as follows:
\begin{prob}[Inverse filtering]
\label{prob:inversion}
Given a filter $P(z)$ which is necessarily not
bi-stable or bi-causal (i.e., $P(z)^{-1}$ can be unstable or noncausal),
find an FIR filter $Q(z)$ which minimizes
\[
\left\| (QP-1)W \right\|_\infty = \max_{\omega\in[0,\pi]} \left|\left(Q(e^{j\omega})P(e^{j\omega})-1\right)W(e^{j\omega})\right|,
\]
where $W$ is a given stable weighting function.
\end{prob}

The procedure to solve this problem is shown in Section \ref{sec:kyp}.

\section{KYP Lemma for $H^\infty$ Design Problems}
\label{sec:kyp}

In this section, we show that the $H^\infty$ design problems given in the previous section
are efficiently solved via {\em semidefinite programming} \cite{BoyVan}.
For this purpose, we first formulate the problems in state-space representation 
reviewed in Section \ref{sec:preliminaries}.
Then we bring in {\em Kalman-Yakubovich-Popov} (KYP) lemma \cite{And67,Ran96,TuqVai98} 
to reduce the problems into semidefinite programming.

\subsection{State-space representation}
The transfer functions $\left(P(z)-Q(z)\right)W(z)$ and $\left(Q(z)P(z)-1\right)W(z)$
in Problems \ref{prob:approximation} and \ref{prob:inversion}, respectively,
can be described in a form of
\begin{equation}
 T(z) = T_1(z) + Q(z)T_2(z),
 \label{eq:Tz}
\end{equation}
where
\[
 T_1(z) = P(z)W(z),\quad T_2(z) = -W(z),
\]
for Problem \ref{prob:approximation} and 
\[
 T_1(z) = -W(z), \quad T_2(z) = P(z)W(z),
\]
for Problem \ref{prob:inversion}.
Therefore, our problems are described by the following min-max optimization:
\begin{equation}
\min_{Q(z)\in\FIR_N} \|T_1+QT_2\|_\infty=\min_{Q(z)\in\FIR_N} \max_{\omega\in[0,\pi]} \left|T_1(e^{j\omega})+Q(e^{j\omega})T_2(e^{j\omega})\right|,
\label{eq:minmax}
\end{equation}
where $\FIR_N$ is the set of $N$-th order FIR filters, that is,
\[
 \FIR_N := \left\{Q(z): Q(z)=\sum_{i=0}^N \alpha_i z^{-i}, \alpha_i\in\real\right\}.
\]
To reduce the problem of minimizing (\ref{eq:minmax}) to semidefinite programming,
we use state-space representation for $T_1(z)$ and $T_2(z)$ in (\ref{eq:Tz}).
Let $\{A_i,B_i, C_i, D_i\}$ $(i=1,2)$ are
state-space matrices of $T_i(z)$ in (\ref{eq:Tz}),
that is,
\[
 T_i(z) = C_i(zI-A_i)^{-1}B_i + D_i =:\abcd{A_i}{B_i}{C_i}{D_i},\quad i=1,2.
\]
Also, a state-space representation of an FIR filter $Q(z)$
is given by
\newcommand{\FIRabcd}{
 \left[
  \begin{array}{ccccc|c}
   0&1&0&\ldots&0&0\\
   0&0&1&\ddots&\vdots&\vdots\\
   0&0&\ddots&\ddots&0&\vdots\\
   \vdots&\vdots&\ddots&0&1&0\\
   0&0&\ldots&0&0&1\\\hline
   \vec{\alpha}_N&\vec{\alpha}_{N-1}&\ldots&\vec{\alpha}_2&\vec{\alpha}_1&\vec{\alpha}_0\\
  \end{array}
 \right]\!(z)
}
\begin{equation}
  Q(z) = \sum_{n=0}^N \vec{\alpha}_nz^{-n}=\FIRabcd=:\abcd{A_q}{B_q}{\vec{\alpha}_{N:1}}{\vec{\alpha}_0}\!(z),
 \label{eq:FIR}
\end{equation}
where
$\vec{\alpha}_{N:1}:=\left[\begin{array}{ccccc}\vec{\alpha}_N&\vec{\alpha}_{N-1}&\ldots&\vec{\alpha}_1\end{array}\right]$.

By using these state-space matrices,
we obtain a state-space representation
of $T(z)$ in (\ref{eq:Tz}) as
\begin{equation}
 T(z) = \left[\begin{array}{ccc|c}
	A_1&0&0&B_1\\
	0&A_2&0&B_2\\
	0&B_qC_2&A_q&B_qD_2\\\hline
	C_1&\vec{\alpha}_0C_2&\vec{\alpha}_{N:1}&D_1+\vec{\alpha}_0D_2
	\end{array}\right]\!(z)
  =:\abcd{A}{B}{C(\vec{\alpha}_{N:0})}{D(\vec{\alpha}_{0})}\!(z).
 \label{eq:Tz-ss}
\end{equation}
Note that the FIR parameters $\vec{\alpha}_0,\vec{\alpha}_1,\ldots,\vec{\alpha}_N$
depend affinely on $C$ and $D$, and are independent of $A$ and $B$.
This property is a key to describe our problem into semidefinite programming.

\subsection{Semidefinite programming by KYP lemma}

The optimization in (\ref{eq:minmax}) can be equivalently described 
by the following minimization problem:
\begin{gather}
 \text{minimize } \gamma \text{ subject to } Q(z)\in\FIR_N \text{ and}\nonumber\\
 \max_{\omega\in[0,\pi]} \left|T_1(e^{j\omega})+Q(e^{j\omega})T_2(e^{j\omega})\right|\leq \gamma.
 \label{eq:gamma}
\end{gather}
To describe this optimization in semidefinite programming,
we adopt the following  lemma \cite{And67,Ran96,TuqVai98}:
\begin{lem}[KYP lemma]
Suppose
\[
 T(z) = \abcd{A}{B}{C}{D}\!(z)
\]
is stable,
and the state-space representation $\{A,B,C,D\}$ of $T(z)$ is minimal%
\footnote{%
 For minimality of state-space representation, see Section \ref{sec:preliminaries} or Chapter 26 in \cite{Rug}.%
}.
Let $\gamma>0$.
Then the following are equivalent conditions:
\begin{enumerate}
\item $\|T\|_\infty \leq \gamma$.
\item There exists a positive definite matrix $X$ such that
\[
 \begin{bmatrix} 
  A^\top XA - X & A^\top XB &C^\top\\
  B^\top XA & B^\top XB-\gamma^2 & D\\
  C & D & -1
 \end{bmatrix}\leq 0.
\]
\end{enumerate}
\end{lem}

By using this lemma, we obtain the following theorem:
\begin{thm}
The inequality (\ref{eq:gamma}) holds if and only if
there exists $X>0$ such that
\begin{equation}
 \begin{bmatrix} 
  A^\top XA - X & A^\top XB & C(\alpha_{N:0})^\top\\
  B^\top XA & B^\top XB-\gamma^2 & D(\alpha_0)\\
  C(\alpha_{N:0}) & D(\alpha_0) & -1
 \end{bmatrix}\leq 0,
 \label{eq:KYP}
\end{equation}
where $A$, $B$, $C(\alpha_{N:0})$, and $D(\alpha_0)$ are
given in (\ref{eq:Tz-ss}).
\end{thm}

By this, the optimal FIR parameters $\vec{\alpha}_0,\vec{\alpha}_1,\ldots,\vec{\alpha}_N$
can be obtained as follows.
Let $\vec{x}$ be the vector consisting of all variables in
$\vec{\alpha}_{N:0}$, $\vec{X}$, and $\vec{\gamma}^2$
in (\ref{eq:KYP}).
The matrix in (\ref{eq:KYP})
is {\em affine}
with respect to these variables,
and hence, can be rewritten in the form
\[
M(\vec{x}) = M_0 + \sum_{i=1}^L M_i\vec{x}_i,
\]
where $M_i$ is a symmetric matrix and $\vec{x}_i$ is the $i$-th 
entry of $\vec{x}$.
Let $v\in\{0,1\}^L$ be a vector such that $v^\top\vec{x}=\vec{\gamma}^2$.
Our problem is then described by semidefinite programming as follows:
\[
 \text{minimize } v^\top \vec{x} \text{ subject to } M(\vec{x})\leq 0.
\]
By this, we can effectively approach the optimal parameters $\vec{\alpha}_{N:0}$
by numerical optimization softwares.
For MATLAB codes of the semidefinite programming above, see Section \ref{sec:matlab}.

\section{Finite Frequency Design of FIR Digital Filters}
\label{sec:prob:finite-frequency}

By the $H^\infty$ design discussed in the previous section,
we can guarantee the maximum gain of the frequency response of 
$T = (P-Q)W$ (approximation) or $T = (QP-1)W$ (inversion)
over the {\em whole frequency range} $[0,\pi]$.
Some applications, however, do not need minimize the gain over the whole range $[0,\pi]$,
but a finite frequency range $\Omega \subset [0,\pi]$.
Design of noise shaping $\Delta\Sigma$ modulators is one example of such requirement \cite{NagYam09}.
In this section, we consider such optimization, called {\em finite frequency optimization}.
We first consider the approximation problem over a finite frequency range.
\begin{prob}[Finite frequency approximation]
\label{prob:approximation2}
Given a filter $P(z)$ and a finite frequency range $\Omega\subset[0, \pi]$,
find an FIR filter $Q(z)$ which minimizes
\[
 V_\Omega(P-Q) := \max_{\omega\in \Omega}\left|P\left(e^{j\omega}\right)-Q\left(e^{j\omega}\right)\right|.
\]
\end{prob}
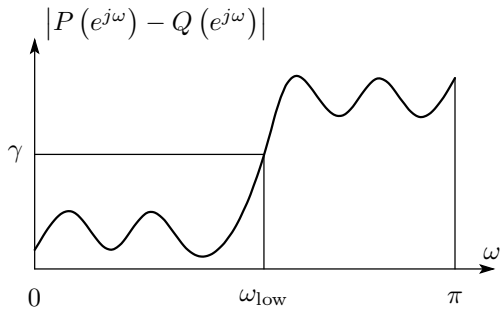
\begin{figure}[tb]
\begin{center}
\unitlength 0.1in
\begin{picture}( 28.6000, 14.3500)( -0.6000,-18.6500)
%
\special{pn 8}%
\special{pa 400 1800}%
\special{pa 400 600}%
\special{fp}%
\special{sh 1}%
\special{pa 400 600}%
\special{pa 380 668}%
\special{pa 400 654}%
\special{pa 420 668}%
\special{pa 400 600}%
\special{fp}%
%
\special{pn 8}%
\special{pa 400 1800}%
\special{pa 2800 1800}%
\special{fp}%
\special{sh 1}%
\special{pa 2800 1800}%
\special{pa 2734 1780}%
\special{pa 2748 1800}%
\special{pa 2734 1820}%
\special{pa 2800 1800}%
\special{fp}%
%
\special{pn 8}%
\special{pa 400 1200}%
\special{pa 1600 1200}%
\special{fp}%
\special{pa 1600 1200}%
\special{pa 1600 1800}%
\special{fp}%
%
\special{pn 13}%
\special{pa 400 1700}%
\special{pa 424 1662}%
\special{pa 446 1624}%
\special{pa 468 1590}%
\special{pa 492 1558}%
\special{pa 514 1532}%
\special{pa 536 1512}%
\special{pa 558 1500}%
\special{pa 582 1496}%
\special{pa 604 1502}%
\special{pa 626 1518}%
\special{pa 650 1544}%
\special{pa 672 1572}%
\special{pa 694 1604}%
\special{pa 718 1636}%
\special{pa 740 1664}%
\special{pa 762 1686}%
\special{pa 786 1698}%
\special{pa 808 1700}%
\special{pa 830 1688}%
\special{pa 854 1668}%
\special{pa 876 1640}%
\special{pa 898 1608}%
\special{pa 920 1576}%
\special{pa 944 1546}%
\special{pa 966 1522}%
\special{pa 988 1504}%
\special{pa 1012 1500}%
\special{pa 1034 1506}%
\special{pa 1056 1522}%
\special{pa 1080 1544}%
\special{pa 1102 1574}%
\special{pa 1124 1604}%
\special{pa 1148 1636}%
\special{pa 1170 1666}%
\special{pa 1192 1692}%
\special{pa 1216 1712}%
\special{pa 1238 1726}%
\special{pa 1260 1734}%
\special{pa 1282 1736}%
\special{pa 1304 1732}%
\special{pa 1328 1722}%
\special{pa 1348 1708}%
\special{pa 1370 1688}%
\special{pa 1392 1666}%
\special{pa 1412 1638}%
\special{pa 1434 1608}%
\special{pa 1454 1574}%
\special{pa 1474 1536}%
\special{pa 1492 1496}%
\special{pa 1512 1454}%
\special{pa 1530 1410}%
\special{pa 1546 1364}%
\special{pa 1564 1316}%
\special{pa 1580 1268}%
\special{pa 1596 1218}%
\special{pa 1610 1170}%
\special{pa 1624 1120}%
\special{pa 1638 1072}%
\special{pa 1650 1024}%
\special{pa 1662 980}%
\special{pa 1676 938}%
\special{pa 1688 900}%
\special{pa 1702 868}%
\special{pa 1714 838}%
\special{pa 1730 816}%
\special{pa 1744 800}%
\special{pa 1760 790}%
\special{pa 1776 788}%
\special{pa 1794 796}%
\special{pa 1814 812}%
\special{pa 1834 836}%
\special{pa 1856 864}%
\special{pa 1878 896}%
\special{pa 1902 928}%
\special{pa 1924 956}%
\special{pa 1948 980}%
\special{pa 1972 996}%
\special{pa 1996 1000}%
\special{pa 2020 994}%
\special{pa 2042 976}%
\special{pa 2066 950}%
\special{pa 2088 918}%
\special{pa 2110 886}%
\special{pa 2134 854}%
\special{pa 2156 828}%
\special{pa 2178 808}%
\special{pa 2200 800}%
\special{pa 2222 806}%
\special{pa 2246 820}%
\special{pa 2268 844}%
\special{pa 2290 874}%
\special{pa 2312 906}%
\special{pa 2336 936}%
\special{pa 2358 966}%
\special{pa 2382 988}%
\special{pa 2404 1002}%
\special{pa 2426 1006}%
\special{pa 2450 998}%
\special{pa 2472 984}%
\special{pa 2494 962}%
\special{pa 2516 934}%
\special{pa 2540 902}%
\special{pa 2562 866}%
\special{pa 2584 828}%
\special{pa 2600 800}%
\special{sp}%
\put(4.5000,-6.0000){\makebox(0,0)[lb]{$\left|P\left(e^{j\omega}\right)-Q\left(e^{j\omega}\right)\right|$}}%
\put(27.5000,-17.5000){\makebox(0,0)[lb]{$\omega$}}%
\put(16.0000,-19.5000){\makebox(0,0){$\omega_\low$}}%
\put(3.0000,-12.0000){\makebox(0,0){$\gamma$}}%
\put(4.0000,-19.5000){\makebox(0,0){$0$}}%
\put(26.0000,-19.5000){\makebox(0,0){$\pi$}}%
%
\special{pn 8}%
\special{pa 2600 1800}%
\special{pa 2600 800}%
\special{fp}%
\end{picture}%
\end{center}
\caption{Finite frequency approximation (Problem \ref{prob:approximation2}):
the error gain $\left|P\left(e^{j\omega}\right)-Q\left(e^{j\omega}\right)\right|$
is minimized over the finite frequency range $\Omega_\low=[0,\omega_\low]$.}
\label{fig:ff-approximation}
\end{figure}
Figure \ref{fig:ff-approximation} illustrates the above problem for a finite frequency range
$\Omega=\Omega_\low=[0,\omega_\low]$, where $\omega_\low\in(0,\pi]$.
We seek an FIR filter which minimizes $V_\Omega(P-Q)$ over the finite frequency range $\Omega$,
and do not care about the other range $[0,\pi]\setminus\Omega$.
We can also formulate the inversion problem over a finite frequency range.
\begin{prob}[Finite frequency inversion]
\label{prob:inversion2}
Given a filter $P(z)$ and a finite frequency range $\Omega\subset[0, \pi]$,
find an FIR filter $Q(z)$ which minimizes
\[
 V_\Omega(QP-1) := \max_{\omega\in \Omega}\left|Q(e^{j\omega})P(e^{j\omega})-1\right|.
\]
\end{prob}

These problems are also fundamental in digital signal processing.
We will show in the next section that these problems can be also described in
semidefinite programming via generalized KYP lemma.

\section{Generalized KYP Lemma for Finite Frequency Design Problems}
\label{sec:gkyp}
In this section, we reduce the problems given in the previous section to semidefinite programming.
As in the $H^\infty$ optimization, we first formulate the problems in state-space representation,
and then derive semidefinite programming via {\em generalized KYP lemma} \cite{IwaHar05}.
\subsection{State-space representation}
As in the $H^\infty$ optimization in Section \ref{sec:kyp},
we employ state-space representation.
Let $T(z)=P(z)-Q(z)$ for the approximation problem or $T(z)=P(z)Q(z)-1$ for the inversion problem.
Then $T(z)$ can be described by
$T(z) = T_1(z)+Q(z)T_2(z)$ as in (\ref{eq:Tz}).
Then our problems are described by the following min-max optimization:
\begin{equation}
\min_{Q(z)\in\FIR_N} V_\Omega\left(T_1+QT_2\right)=\min_{Q(z)\in\FIR_N} \max_{\omega\in\Omega} \left|T_1(e^{j\omega})+Q(e^{j\omega})T_2(e^{j\omega})\right|.
\label{eq:minmax-ff}
\end{equation}

Let $\{A_i,B_i,C_i,D_i\}$, $i=1,2$ be state-space matrices of $T_i(z)$.
By using the same technique as in Section \ref{sec:kyp}, we can obtain
a state-space representation of $T(z)$ as
\begin{equation}
 T(z) = \abcd{A}{B}{C(\vec{\alpha}_{N:0})}{D(\vec{\alpha}_{0})}\!(z),
 \label{eq:Tz-ff}
\end{equation}
where $\alpha_{N:0}=[\alpha_N,\ldots,\alpha_0]$ is the coefficient vector of the FIR filter to be designed
as defined in (\ref{eq:FIR}).
\subsection{Semidefinite programming by generalized KYP lemma}

The optimization in (\ref{eq:minmax-ff}) can be equivalently described
by the following problem:
\begin{gather}
 \text{minimize } \gamma \text{ subject to } Q(z)\in\FIR_N \text{ and}\nonumber\\
 \max_{\omega\in[0,\pi]} \left|T_1(e^{j\omega})+Q(e^{j\omega})T_2(e^{j\omega})\right|\leq \gamma
 \label{eq:gamma-ff}
\end{gather}
To describe this optimization in semidefinite programming,
we adopt the following  lemma \cite{IwaHar05}:

\begin{lem}[Generalized KYP Lemma]
Suppose
\[
 T(z) = \abcd{A}{B}{C}{D}\!(z)
\]
is stable,
and the state-space representation $\{A,B,C,D\}$ of $T(z)$ is minimal.
Let $\Omega$ be a closed interval $[\omega_1,\omega_2] \subset [0,\pi]$.
Let $\gamma>0$.
Then the following are equivalent conditions:
\begin{enumerate}
\item $V_\Omega(T)=\max_{\omega\in[\omega_1,\omega_2]}\left|T\left(e^{j\omega}\right)\right|\leq \gamma$.
\item
There exist symmetric matrices $Y>0$ and $X$ such that
\[
\left[
\begin{array}{ccc}
M_{1}(X,Y) & M_{2}(X,Y) & C^\top\\
\overline{M}_{2}(X,Y)^\top & M_{3}(X,\gamma^2) & D\\
C & D & -1\\
\end{array}
\right]\leq 0,
\]
where
\begin{equation}
\begin{split}
 M_{1}(X,Y)&= A^\top XA + YAe^{-j\omega_0} + A^\top Ye^{j\omega_0} - X - 2Y \cos r,\\
 M_{2}(X,Y) &= A^\top XB + YBe^{-j\omega_0},\quad
 \overline{M}_{2}(X,Y) = A^\top XB + YBe^{j\omega_0},\\
 M_{3}(X,\gamma^2) &= B^\top XB - \gamma^2,\quad
 \omega_0 = \frac{\omega_1+\omega_2}{2},\quad r=\frac{\omega_2-\omega_1}{2}.
\end{split}
\label{eq:Ms}
\end{equation}
\end{enumerate}
\end{lem}

By using this lemma, we obtain the following theorem:
\begin{thm}
The inequality (\ref{eq:gamma-ff}) holds if and only if
there exist symmetric matrices $Y>0$ and $X$ such that
\[
\left[
\begin{array}{ccc}
M_{1}(X,Y) & M_{2}(X,Y) & C(\alpha_{N:0})^\top\\
\overline{M}_{2}(X,Y)^\top & M_{3}(X,\gamma^2) & D({\alpha_0})\\
C(\alpha_{N:0}) & D(\alpha_0) & -1\\
\end{array}
\right]\leq 0,
\]
where $M_{1}$, $M_{2}$, and $M_{3}$ are given in (\ref{eq:Ms}),
$A$, $B$, $C(\alpha_{N:0})$, and $D(\alpha_0)$ are given in (\ref{eq:Tz-ff}).
\end{thm}

By this theorem, we can obtain the coefficients $\alpha_0,\ldots,\alpha_N$ of the optimal FIR filter 
by semidefinite programming as mentioned in Section \ref{sec:kyp}.
MATLAB codes for the semidefinite programming are shown in Section \ref{sec:matlab}.

\section{MATLAB Codes for Semidefinite Programming}
\label{sec:matlab}

In this section, we give MATLAB codes for the semidefinite programming derived 
in previous sections.
Note that the MATLAB codes for solving {\bf Problems 1} to {\bf 4}
are also available at the following web site:
\begin{center}
\verb= http://www-ics.acs.i.kyoto-u.ac.jp/~nagahara/fir/=
\end{center}
Note also that to execute the codes in this section, 
Control System Toolbox \cite{CT}, YALMIP \cite{YALMIP}, and SeDuMi \cite{SeDuMi} are needed.
YALMIP and SeDuMi are free softwares for solving optimization problems including semidefinite programming
which is treated in this article.
\subsection{FIR approximation of IIR filters by $H^\infty$ norm}
\begin{verbatim}
function [q,gmin] = approxFIRhinf(P,W,N);
% [q,gmin]=approxFIRhinf(P,W) computes the
% H-infinity optimal approximated FIR filter Q(z) which minimizes
%   J(Q) = ||(P-Q)W||,
% the maximum frequency gain of (P-Q)W.
% This design uses SDP via the KYP lemma.
% 
% Inputs:
%   P: Target stable linear system in SS object
%   W: Weighting stable linear system in SS object
%   N: Order of the FIR filter to be designed
%
% Outputs:
%   q: The optimal FIR filter coefficients
%   gmin: The optimal value
%

%% Initialization
T1 = P*W;
T2 = -W;
[A1,B1,C1,D1]=ssdata(T1);
[A2,B2,C2,D2]=ssdata(T2);
n1 = size(A1,1);
n2 = size(A2,1);

%% FIR filter to be designed
Aq = circshift(eye(N),-1);
Aq(N,1) = 0;
Bq = [zeros(N-1,1);1];

%% Semidefinite Programming
A = [A1, zeros(n1,n2), zeros(n1,N);
     zeros(n2,n1), A2, zeros(n2,N);
     zeros(N,n1),Bq*C2, Aq];
B = [B1;B2;Bq*D2];

NN = size(A,1);

X = sdpvar(NN,NN,'symmetric');
alpha_N1 = sdpvar(1,N);
alpha_0 = sdpvar(1,1);
gamma = sdpvar(1,1);

M1 = A'*X*A-X;
M2 = A'*X*B;
M3 = B'*X*B-gamma;

C = [C1, alpha_0*C2, alpha_N1];
D = D1 + alpha_0*D2;

M = [M1, M2, C'; M2', M3, D; C, D, -gamma];

F = set(M < 0) + set(X > 0) + set(gamma > 0);

solvesdp(F,gamma);

%% Optimal FIR filter coefficients
q = fliplr([double(alpha_N1),double(alpha_0)]);
gmin = double(gamma);
\end{verbatim}
\subsection{Inverse FIR filtering by $H^\infty$ norm}
\begin{verbatim}
function [q,gmin] = inverseFIRhinf(P,W,N,n);
% [q,gmin]=inverseFIRhinf(P,W,N,n) computes the
% H-infinity optimal (delayed) inverse FIR filter Q(z) which minimizes
%   J(Q) = ||(QP-z^(-n))W||,
% the maximum frequency gain of (QP-z^(-n))W.
% This design uses SDP via the KYP lemma.
% 
% Inputs:
%   P: Target stable linear system in SS object
%   W: Weighting stable linear system in SS object
%   N: Order of the FIR filter to be designed
%   n: Delay (this can be omitted; default value=0);
%
% Outputs:
%   q: The optimal FIR filter coefficients
%   gmin: The optimal value
%

if nargin==3
    n=0
end

%% Initialization
z = tf('z');
T1 = -z^(-n)*W;
T2 = P*W;
[A1,B1,C1,D1]=ssdata(T1);
[A2,B2,C2,D2]=ssdata(T2);
n1 = size(A1,1);
n2 = size(A2,1);

%% FIR filter to be designed
Aq = circshift(eye(N),-1);
Aq(N,1) = 0;
Bq = [zeros(N-1,1);1];

%% Semidefinite Programming
A = [A1, zeros(n1,n2), zeros(n1,N);
     zeros(n2,n1), A2, zeros(n2,N);
     zeros(N,n1),Bq*C2, Aq];
B = [B1;B2;Bq*D2];

NN = size(A,1);

X = sdpvar(NN,NN,'symmetric');
alpha_N1 = sdpvar(1,N);
alpha_0 = sdpvar(1,1);
gamma = sdpvar(1,1);

M1 = A'*X*A-X;
M2 = A'*X*B;
M3 = B'*X*B-gamma;

C = [C1, alpha_0*C2, alpha_N1];
D = D1 + alpha_0*D2;

M = [M1, M2, C'; M2', M3, D; C, D, -gamma];

F = set(M < 0) + set(X > 0) + set(gamma > 0);

solvesdp(F,gamma);

%% Optimal FIR filter coefficients
q = fliplr([double(alpha_N1),double(alpha_0)]);
gmin = double(gamma);
\end{verbatim}
\subsection{FIR approximation of IIR filters by finite-frequency min-max}
\begin{verbatim}
function [q,gmin] = approxFIRff(P,Omega,N);
% [q,gmin]=approxFIRff(P,Omega,N) computes the
% Finite-frequency optimal approximated FIR filter Q(z) which minimizes
%   J(Q) = max{|P(exp(jw))-Q(exp(jw))|, w in Omega}l.
% the maximum frequency gain of P-Q in a frequency band Omega.
% This design uses SDP via the generalized KYP lemma.
% 
% Inputs:
%   P: Target stable linear system in SS object
%   Omega: Frequency band in 1x2 vector [w1,w2]
%   N: Order of the FIR filter to be designed
%
% Outputs:
%   q: The optimal FIR filter coefficients
%   gmin: The optimal value
%

%% Initialization
[A1,B1,C1,D1]=ssdata(P);
n1 = size(A1,1);

%% FIR filter to be designed
Aq = circshift(eye(N),-1);
Aq(N,1) = 0;
Bq = [zeros(N-1,1);1];

%% Semidefinite Programming
A = blkdiag(A1,Aq);
B = [B1;-Bq];

NN = size(A,1);

omega0 = (Omega(1)+Omega(2))/2;
omegab = (Omega(2)-Omega(1))/2;

P = sdpvar(NN,NN,'symmetric');
Q = sdpvar(NN,NN,'symmetric');
alpha_N1 = sdpvar(1,N);
alpha_0 = sdpvar(1,1);
g = sdpvar(1,1);

C = [C1, alpha_N1];
D = D1 - alpha_0;

M1r = A'*P*A+Q*A*cos(omega0)+A'*Q*cos(omega0)-P-2*Q*cos(omegab);
M2r = A'*P*B + Q*B*cos(omega0);
M3r = B'*P*B-g;
M1i = A'*Q*sin(omega0)-Q*A*sin(omega0);
M21i = -Q*B*sin(omega0);
M22i = B'*Q*sin(omega0);
Mr = [M1r,M2r,C';M2r',M3r,D;C,D,-1];
Mi = [M1i, M21i, zeros(NN,1);M22i, 0, 0; zeros(1,NN),0,0];
M = [Mr, Mi; -Mi, Mr];

F = set(M < 0) + set(Q > 0) + set(g > 0);

solvesdp(F,g);

%% Optimal FIR filter coefficients
q = fliplr([double(alpha_N1),double(alpha_0)]);
gmin = double(g);
\end{verbatim}
\subsection{Inverse FIR filtering by finite-frequency min-max}
\begin{verbatim}
function [q,gmin] = inverseFIRff(P,Omega,N,n);
% [q,gmin]=inverseFIRff(P,Omega,N,n) computes the
% Finite-frequency optimal (delayed) inverse FIR filter Q(z) which minimizes
%   J(Q) = max{|Q(exp(jw)P(exp(jw))-exp(-jwn)|, w in Omega}.
% the maximum frequency gain of QP-z^(-n) in a frequency band Omega.
% This design uses SDP via the generalized KYP lemma.
% 
% Inputs:
%   P: Target stable linear system in SS object
%   Omega: Frequency band in 1x2 vector [w1,w2]
%   N: Order of the FIR filter to be designed
%   n: Delay (this can be omitted; default value=0);
%
% Outputs:
%   q: The optimal FIR filter coefficients
%   gmin: The optimal value
%

if nargin==3
    n=0
end

%% Initialization
z = tf('z');
T1 = -z^(-n);
T2 = P;
[A1,B1,C1,D1]=ssdata(T1);
[A2,B2,C2,D2]=ssdata(T2);
n1 = size(A1,1);
n2 = size(A2,1);

%% FIR filter to be designed
Aq = circshift(eye(N),-1);
Aq(N,1) = 0;
Bq = [zeros(N-1,1);1];

%% Semidefinite Programming
A = [A1, zeros(n1,n2), zeros(n1,N);
     zeros(n2,n1), A2, zeros(n2,N);
     zeros(N,n1),Bq*C2, Aq];
B = [B1;B2;Bq*D2];

NN = size(A,1);

omega0 = (Omega(1)+Omega(2))/2;
omegab = (Omega(2)-Omega(1))/2;

P = sdpvar(NN,NN,'symmetric');
Q = sdpvar(NN,NN,'symmetric');
alpha_N1 = sdpvar(1,N);
alpha_0 = sdpvar(1,1);
g = sdpvar(1,1);

C = [C1, alpha_0*C2, alpha_N1];
D = D1 + alpha_0*D2;

M1r = A'*P*A+Q*A*cos(omega0)+A'*Q*cos(omega0)-P-2*Q*cos(omegab);
M2r = A'*P*B + Q*B*cos(omega0);
M3r = B'*P*B-g;
M1i = A'*Q*sin(omega0)-Q*A*sin(omega0);
M21i = -Q*B*sin(omega0);
M22i = B'*Q*sin(omega0);
Mr = [M1r,M2r,C';M2r',M3r,D;C,D,-1];
Mi = [M1i, M21i, zeros(NN,1);M22i, 0, 0; zeros(1,NN),0,0];
M = [Mr, Mi; -Mi, Mr];

F = set(M < 0) + set(Q > 0) + set(g > 0);

solvesdp(F,g);

%% Optimal FIR filter coefficients
q = fliplr([double(alpha_N1),double(alpha_0)]);
gmin = double(g);
\end{verbatim}

\section{Examples}
\begin{figure}[tbp]
\begin{center}
\includegraphics[width=0.7\linewidth]{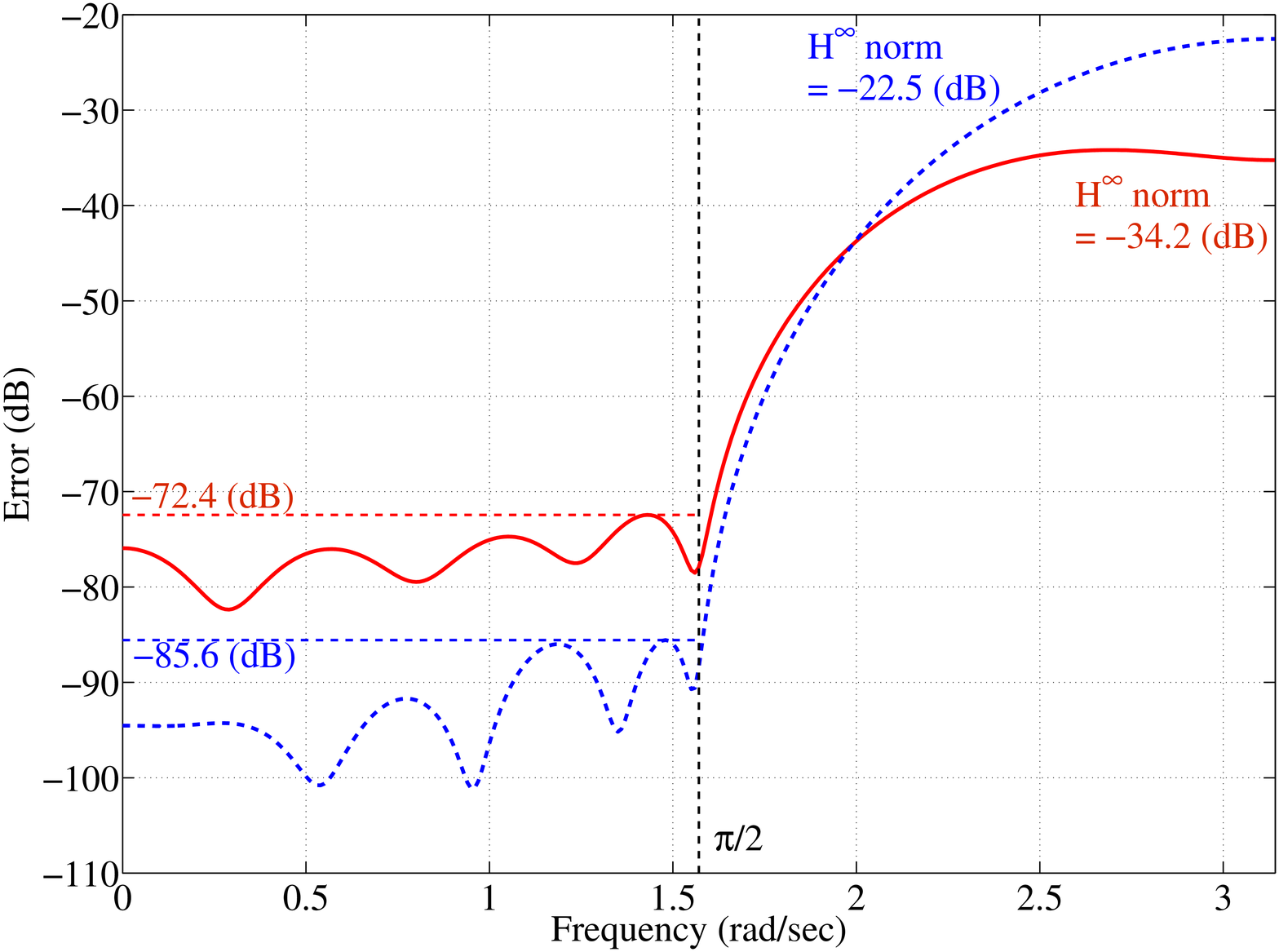}
\end{center}
\caption{The gain of the error $E(z)=P(z)-Q(z)$ for $H^\infty$ optimization (solid)
and finite-frequency min-max optimization (dash)}
\label{fig:error}
\end{figure}
By the MATLAB codes given in the previous section, 
we design FIR filters
for {\bf Problems 1} and {\bf 3}.
Let the FIR filter order $N=8$.
The target filter is the second order lowpass Butterworth
filter with cutoff frequency $\pi/2$.
This can be computed by \verb=butter(2,1/2)= in MATLAB.
The weighting transfer function in {\bf Problem 1} is chosen
by a 8th order lowpass Chebyshev filter,
computed by \verb=cheby1(8,1/2,1/2)= in MATLAB.
The frequency band for {\bf Problem 3} is $\Omega=[0,\pi/2]$.
Figure \ref{fig:error} shows the gain of the error $E(z):=P(z)-Q(z)$.
We can see that the $H^\infty$ optimal filter (the solution of {\bf Problem 1}),
say $Q_1(z)$,
shows the lower $H^\infty$ norm than the finite-frequency min-max design
(the solution of {\bf Problem 3}), say $Q_2(z)$.
On the other hand, in the frequency band $[0,\pi/2]$,
$Q_1(z)$ shows the larger error than $Q_2(z)$.

\section{Conclusion}
In this article, we consider four problems,
FIR approximation and inverse FIR filtering of IIR filters by $H^\infty$ and finite-frequency min-max,
which are fundamental in signal processing.
By using KYP and generalized KYP lemmas, the problems are all solvable
via semidefinite programming.
We show MATLAB codes for the programming, and show examples of designing FIR filters.


\end{document}